\documentclass[12pt]{article}

\pdfoutput=1

\usepackage[T1]{fontenc}
\usepackage[margin=1in]{geometry}
\usepackage{graphicx}
\usepackage{amsmath,amssymb}
\usepackage{natbib}
\usepackage{authblk}
\usepackage{hyperref}
\usepackage{booktabs}
\usepackage{tabularx}
\usepackage{changepage}
\usepackage{caption}
\usepackage{times}
\usepackage{float}
\hypersetup{colorlinks=true, linkcolor=blue, citecolor=blue, urlcolor=blue}

\title{Dark Matter in the Galaxy: Insights from the Dynamics of Stellar Streams}

\author[1]{Mohammad Ayaz}
\author[1,2]{Priya Hasan}
\affil[1]{Email: mohdayaz151@gmail.com, priya.hasan@gmail.com\\

Department of Physics, Maulana Azad National Urdu University, Telecom Nagar, Gachibowli, Hyderabad 500032, Telangana, India}
\affil[2]{Inter-University Centre for Astronomy and Astrophysics, Post Bag 4, Ganeshkhind, Savitribai Phule Pune University Campus, Pune 411007, Maharashtra, India}

\date{}

\begin{document}

\maketitle

\begin{sloppypar}
\vspace{-1.5em}
\begin{abstract}
\noindent Dark matter makes up most of the mass in galaxies and galaxy clusters. Despite making up most of the matter in the Universe, it has never been directly detected and its fundamental nature and distribution remain unknown. Instead, astronomers search for evidence of dark matter through indirect signatures of its behavior in space, including unexplained signals observed near the center of our galaxy, the Milky Way. Although rotation curves, gravitational lensing, and the cosmic microwave background point to a large, gravitationally dominant invisible mass, transforming these astrophysical constraints into a precise model of the Milky Way's dark matter halo that accurately resolves its geometry, smoothness, and fine-scale features remains an outstanding problem in galactic dynamics. Addressing this challenge requires tracers sensitive to the gravitational potential on galactic scales. In this review we examine what stellar streams (the tidal debris of disrupted globular clusters and dwarf galaxies) can reveal about the Milky Way's dark matter halo. After summarizing the broader evidence for and candidates of dark matter, we describe the theoretical basis for using streams as halo tracers. We then detail the observational signatures, such as coherence, gaps, spurs, and large-scale wiggles, that connect stream morphology to halo properties. Five streams, namely Sagittarius, GD-1, Palomar 5, Orphan-Chenab, and Cetus Polar, serve as case studies spanning a range of orbital geometries and halo regimes. We then describe the methodological frameworks, from analytic orbit fitting to statistical and Bayesian inference, used to interpret these signatures, and compare idealized with realistic halo models. We also outline the principal challenges, including parameter degeneracies, incomplete observational coverage, progenitor sensitivity, and time-dependent perturbations such as the infall of the Large Magellanic Cloud. These factors currently limit the reliability of models of the Galaxy's dark matter distribution. We conclude by discussing how new astrometric and photometric surveys, combined with improved statistical frameworks, may allow stream substructure to discriminate between cold, warm, and self-interacting dark matter. These efforts may help resolve the particle nature of dark matter.
\par\vspace{0.8em}

\end{abstract}
\end{sloppypar}

\section{Introduction}
\begin{figure}[htbp]
\centering
\includegraphics[width=0.75\textwidth]{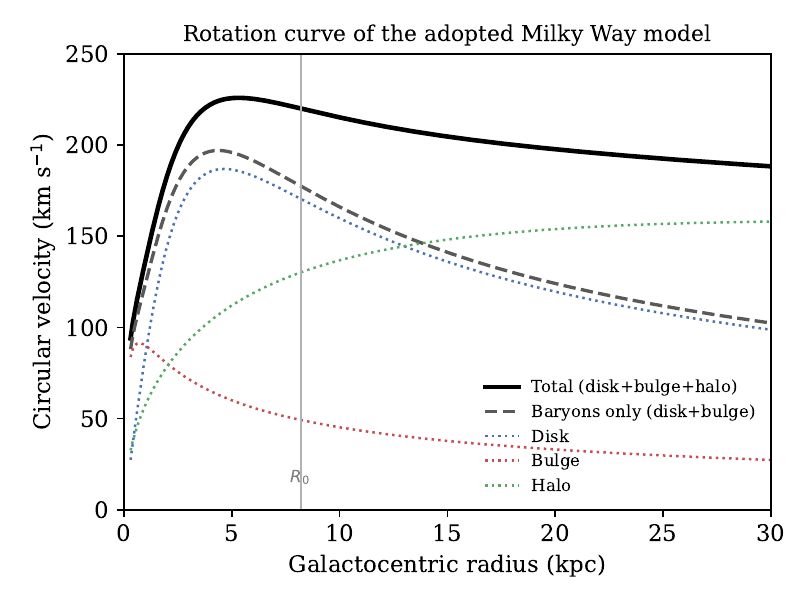}
\caption{Rotation curve of the axisymmetric Milky Way mass model adopted
in this review: a Miyamoto-Nagai disk, a Hernquist bulge, and an NFW halo
\citep{nfw1996, nfw1997}, combined following the component-normalization
scheme of \citet{bovy2015galpy} and calibrated to a circular velocity of
$220\,\mathrm{km\,s^{-1}}$ at the solar radius $R_0 = 8.2\,\mathrm{kpc}$
\citep[structural parameters broadly consistent with][]{bland-hawthorn2016}.
This three-component mass model is computed from closed-form expressions for each potential rather than a direct fit to any individual galaxy data points, acting as the reference model for the
test-particle stream simulations discussed later in this review.}
\label{fig:rotcurve}
\end{figure}

\begin{sloppypar}
The Milky Way is surrounded by a massive dark matter halo, which is responsible for shaping its gravitational potential. Even with rapid advancements in observational techniques and ongoing efforts, determining the local density distribution, substructure details, and properties of the Milky Way halo (the dominant source of the Galactic potential) remains a primary objective for accurately mapping how mass is distributed across the Milky Way and tracing the structural evolution of our Galaxy. The evidence
that most of the matter in the Universe is non-luminous traces back to
\citet{zwicky1933}, who noted that the velocity dispersion of galaxies in
the Coma cluster implied far more mass than could be accounted for by
their luminous content. Four decades later, roughly flat rotation curves in the
Milky Way and other spiral galaxies \citep{rubinford1970,
rubinfordthonnard1980, sofuerubin2001, bland-hawthorn2016} turned this into a galaxy-scale problem, and
gravitational-lensing observations of merging clusters \citep{clowe2006}
later provided some of the most direct evidence that this missing mass is
real, extended, and largely collisionless. The statistics of the cosmic
microwave background independently require a non-baryonic dark matter
component to explain the observed pattern of fluctuations and the
subsequent growth of large-scale structure \citep{planck2018}. Taken
together, this body of evidence leaves little doubt that dark matter
exists in some form; what remains far less settled is what it is made of,
and how it is arranged within an individual galaxy like our own. Both
questions are the subject of this review, though our emphasis is on the
second: using the Milky Way itself, and one particular class of dynamical
tracer within it, as a laboratory for probing the dark matter that
surrounds us.
\end{sloppypar}
Mapping the spatial distribution of the halo is equally important for particle physics. This is because the geometry and shape we assume for the halo govern what signals we expect to detect, and because the
small-scale structure of the halo is itself sensitive to the underlying nature of the dark matter particle. Standard analytic approaches have long
modeled the Milky Way's halo as smooth, spherical, and static
\citep{nfw1996, nfw1997}, an assumption that is mathematically convenient
(Figure~\ref{fig:rotcurve}) but increasingly hard to reconcile with
theory. Cosmological simulations carried out within the $\Lambda$CDM
paradigm instead predict halos that are triaxial, clumpy, and still
evolving today, built up through the continual accretion and disruption
of smaller satellite systems \citep{bullockjohnston2005, johnston2008}. Rotation curves are sensitive only to the roughly spherically averaged mass enclosed within a given radius. Consequently, a halo model calibrated on these observations cannot by itself distinguish between these two pictures, nor can it detect the low-mass subhalos that different dark matter candidates predict in varying abundances. Testing which picture actually
describes the Milky Way, and probing its small-scale content, requires a
tracer that responds to the gravitational field over a much larger
volume, and in more than one dimension.

\enlargethispage{-0.5\baselineskip}
Stellar streams offer an observational solution to this problem. These elongated structures form when
globular clusters or dwarf galaxies are tidally disrupted by the Galactic
potential (Figure~\ref{fig:streamformation}), spreading their stars along
a common orbit that can extend for tens of kiloparsecs
\citep{ibata1994, johnstonhernquistbolte1996, helmiwhite1999}. Because
every star in a stream shares, to first approximation, the same orbital
history, a stream behaves almost like a test particle drawn out into a
long, luminous thread. Such irregularities (where the thread kinks, thins, frays, or splits) trace where the underlying potential, or a passing
clump of dark matter, has done something an idealized spherical halo
would not \citep{ibata2002substructure, koposov2010, bonaca2019}. This sensitivity has
only become useful at scale in the last decade, once the \textit{Gaia}
mission \citep{gaia2016, gaiadr2brown2018, gaiadr3} began providing proper
motions and parallaxes precise enough to trace faint, low-surface-brightness
streams across the whole sky rather than in disconnected, individually targeted fields. In this review we deliberately consider several such
streams together, rather than concentrating on any single well-studied
case, because it is the pattern across systems of different progenitor
type, orbit, and Galactic environment, rather than any single stream in
isolation, that ultimately constrains what the Milky Way's dark matter
halo is actually shaped like, and what that shape and substructure can
tell us about the dark matter itself.

\section{Background}

\subsection{The Dark Matter Problem: Evidence and Candidates}
\label{sec:dmproblem}
\begin{sloppypar}
Our primary focus is on stellar streams and dark matter halos, but it is important to get a clear picture of why astronomers are confident that dark matter exists. This provides essential context,~as the halo-tracing problem discussed in this review is only interesting because of this broader framework. The observational case is based on several largely independent lines of evidence. Flat rotation curves show that the
enclosed mass of spiral galaxies continues to grow well beyond the extent
of their visible disks \citep{rubinford1970, rubinfordthonnard1980}.
Gravitational lensing around merging galaxy clusters (most famously the Bullet Cluster) shows that most of the mass of the cluster is spatially
offset from its hot, X-ray-emitting gas, which constitutes the dominant baryonic component. This confirms that the missing mass is not just hidden baryonic
matter but a distinct, largely collisionless component \citep{clowe2006}.
At cosmological scales, the detailed pattern of temperature fluctuations (the CMB anisotropies) 
in the cosmic microwave background, and the subsequent growth of
large-scale structure from those fluctuations, can only be reproduced in models that include a non-baryonic dark matter component making up
roughly 85\% of the Universe's total matter density
\citep{planck2018}. These three observational probes cover galactic,
cluster, and cosmological scales respectively, and the way these independent measurements support one another establishes the foundation for the broader acceptance of the dark matter hypothesis within the standard $\Lambda$CDM cosmological model.
\end{sloppypar}
\begin{sloppypar}
The leading particle candidates for dark matter span a wide range of
masses and interaction strengths, and the review of \citet{bertone2005}
remains a comprehensive starting point for this space of candidates. Weakly
interacting massive particles (WIMPs), motivated by supersymmetric
extensions of the Standard Model, were for decades the best-motivated
candidate, predicting a roughly universal ``cold'' dark matter behavior on
galactic scales: particles that were non-relativistic when they decoupled
from the early Universe, clustering readily into the hierarchical,
substructure-rich halos that N-body simulations predict. Axions are light pseudoscalar particles, originally proposed to solve the
strong-CP problem in quantum chromodynamics. The energy stored in the
oscillations of the classical axion field behaves as cold dark matter
\citep{preskillwisewilczek1983}. Much lighter axion-like particles, with masses near
$10^{-22}$\,eV, can behave as a wave with a de Broglie wavelength of about
a kiloparsec, an idea known as fuzzy dark matter \citep{hui2017}.
Sterile neutrinos, hypothetical right-handed partners to the known
neutrino flavors, would instead be ``warm'' dark matter: relativistic for
longer in the early Universe, which suppresses the formation of the
smallest-scale structure relative to the cold dark matter case
\citep{dodelsonwidrow1994}. Self-interacting dark matter, in which
particles scatter off one another via a new force, sits somewhere between
these pictures dynamically, potentially altering the inner density
profiles of halos without much changing their large-scale clustering.
Not every viable explanation for the observed dynamics requires a new
particle at all: modified Newtonian dynamics (MOND) proposes instead that
the law of gravity itself departs from the Newtonian form below a
characteristic acceleration scale \citep{milgrom1983}, and while it
remains an alternative framework and struggles to match cluster-scale and
cosmological evidence without additional dark matter, it is a useful
reminder that the case for particle dark matter, however strong, rests on
gravitational inference rather than direct detection.

On these small scales, cold, warm, and self-interacting
dark matter behave quite differently, resulting in distinct predictions for the small-scale
structure of a galaxy's halo, specifically regarding the abundance, density
profile, and minimum mass of the subhalos that should populate it. This is precisely the kind of small-scale structure that direct-detection experiments and gamma-ray searches for dark matter annihilation are
largely insensitive to. However, dynamical tracers within the Milky Way, such as stellar streams, are well positioned to probe these features. The
remainder of this review focuses on how stellar streams function as dynamical probes. We examine what these streams are, which morphological features carry dynamical information, and what they have revealed about the shape, mass, and substructure of the Galaxy's dark matter halo.
\end{sloppypar}

\subsection{Dark Matter Halos in Galactic Dynamics}
\begin{sloppypar}
The missing mass problem leads scientists to conclude that non-baryonic matter (dark matter) exists, because we need something to fill the gap between observed and expected motions of stars in the Galaxy. We call it dark not because of its color, but because of its nature, as it does not interact with light. The dark matter halo concept arises directly from this, derived from the rotation-curve evidence discussed at the beginning of the introduction, which is the discrepancy between the observed motions of stars and gas in galaxies and the gravitational influence expected only from baryonic matter (everything we can see or observe). In the Milky Way and other spiral galaxies, rotation curves do not fall off in the Keplerian way expected from the visible mass alone \citep{rubinford1970,
rubinfordthonnard1980, sofuerubin2001, bland-hawthorn2016}. This observational evidence confirms the existence of an extended, invisible mass component
that continues well beyond the edge of the visible disk. Early models assumed this component to be smooth, spherical, and static, adopting
profiles such as the singular isothermal sphere or the
Navarro-Frenk-White (NFW) profile \citep{nfw1996, nfw1997}, both of which remain widely used due to the simplicity they bring to orbit and potential calculations. Cosmological simulations within $\Lambda$CDM instead predict halos that
are triaxial, clumpy, and evolving, shaped by ongoing accretion
\citep{bullockjohnston2005}. This is because a halo assembled hierarchically out of many
smaller progenitors retains signatures of that assembly history in its
shape, its degree of substructure, and its response to the continued
infall of satellites such as the Large Magellanic Cloud. Determining which
of these pictures, smooth and simple versus clumpy and evolving, more
accurately describes the actual Milky Way is not a matter of theoretical preference. It directly determines what kind of dark matter is consistent with observational data, and it is the main motivation for using dynamical
tracers, and stellar streams in particular, that this review develops in
the following sections.
\end{sloppypar}

\subsection{Formation of Stellar Streams}

\begin{figure}[htbp]
\centering
\includegraphics[width=0.95\textwidth]{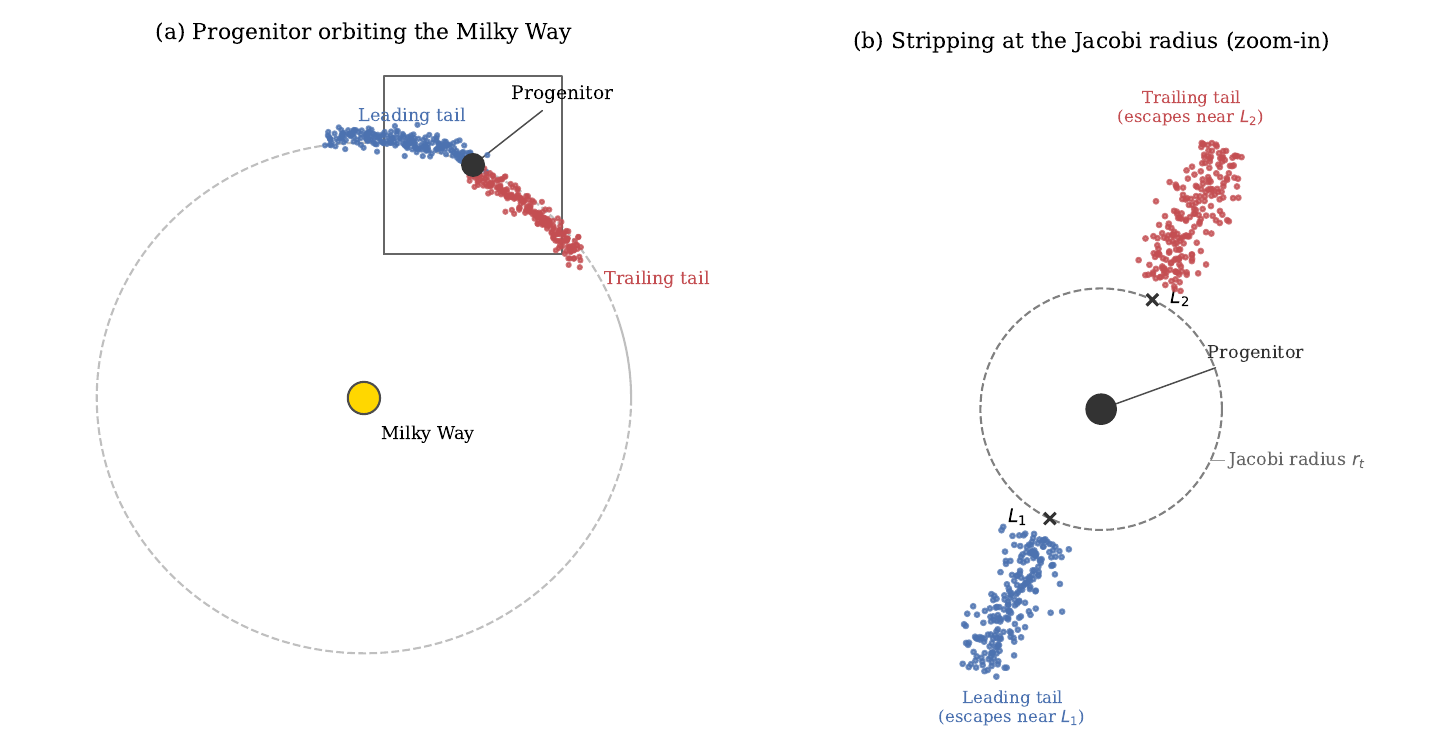}
\caption{Schematic representation of tidal stream formation. Part \text{(a)} shows a progenitor system orbiting the
Milky Way, where tidal forces strip stars from its leading and trailing edges,
forming two tails that trace the progenitor's orbit and evolve into an
extended stellar stream. Part \text{(b)} gives a zoomed-in view of the stripping process where stars become unbound once they cross the Jacobi radius
$r_t$ (Equation~\ref{eq:tidalradius}), escaping preferentially near the
inner Lagrange point $L_1$ (forming the leading tail) and the outer
Lagrange point $L_2$ (forming the trailing tail). The Jacobi surface is drawn here as a circle for clarity, although in reality it is more complex and forms an elongated, teardrop-shaped boundary stretched along the line connecting the progenitor and the Galactic center.}
\label{fig:streamformation}
\end{figure}

Stellar streams form when bound stellar systems, such as globular clusters or
dwarf galaxies, are gradually disrupted by the Milky Way's tidal field
\citep{ibata1994, johnstonhernquistbolte1996}. As a progenitor orbits the
Galaxy, stars near its inner and outer Lagrange points are no longer bound
tightly enough to the progenitor's own gravity to resist the Milky Way's
tidal pull; once stripped, they drift away from the progenitor at a slow,
steady rate, tracing out two tails that extend ahead of and behind the
progenitor along its orbit (Figure~\ref{fig:streamformation}). Stars lost
near the inner Lagrange point tend to fall onto slightly lower-energy,
faster orbits and form what is conventionally called the leading tail,
while stars lost near the outer Lagrange point fall onto slightly
higher-energy, slower orbits and form the trailing tail. Because
disruption proceeds slowly and continuously along a progenitor's orbit,
rather than in a single violent event, a stream typically comprises stars
stripped over many orbital periods, spanning a considerable range of
tidal-stripping epochs by the time it is observed. Even with such a large spatial extent (some streams span more than a hundred
degrees on the sky), streams preserve dynamical signatures of the
gravitational environment in which they are formed \citep{helmiwhite1999}. This is 
because each orbit of the stars in the streams is governed by the same underlying
potential as the progenitor's. This makes streams not just space debris or remnant structures, but extended stellar systems that possess the properties required to function as halo tracers. An individual orbit has fixed values of energy and angular momentum, whereas a fully developed stream traces a sequence of nearby orbits. As a result, the stream maps an entire manifold of adjacent orbits, allowing its observed path to constrain the shape of the potential across galactic scales rather than at a single point.

\begin{sloppypar}
\subsection{Streams as Probes of the Galactic Halo}

Stellar streams provide a direct probe of the dark matter halo, because they respond directly to its gravitational field. The dynamical
evolution of a stream is governed by the collisionless Boltzmann equation,

\begin{equation}
    \frac{df}{dt} = \frac{\partial f}{\partial t}
    + \mathbf{v}\cdot\nabla f - \nabla\Phi\cdot\nabla_{v}f = 0,
    \label{eq:cbe}
\end{equation}

\noindent where $f(\mathbf{r},\mathbf{v},t)$ is the phase-space
distribution function and $\Phi(\mathbf{r},t)$ is the Galactic potential
\citep{binneytremaine2008}. Any lack of smoothness (inhomogeneities) in $\Phi$ naturally distorts $f$, leading to observational variations in stream structures.

\noindent Taking velocity moments of Equation~\ref{eq:cbe} for a steady-state system
leads to the Jeans equations, which relate the potential to observable
velocity dispersions:

\begin{equation}
    \frac{\partial(\rho\langle v_i v_j\rangle)}{\partial x_j}
    + \rho\,\frac{\partial\Phi}{\partial x_i} = 0,
    \label{eq:jeans}
\end{equation}

\noindent where $\rho$ is the stellar density and $\langle v_i
v_j\rangle$ are the velocity moments. These equations assume a steady state, which a stream reaches only once
it has phase-mixed, so they relate the potential to velocity dispersions
only for such relaxed systems. Structural heating and scattering lead to observable gaps and asymmetries in a stream, revealing halo triaxiality, anisotropy, or substructure \citep{helmi2004}.
\end{sloppypar}
\begin{sloppypar}
Stream thickness and continuity are further governed by the tidal radius
of the progenitor system,

\begin{equation}
    r_t \approx R\left(\frac{m}{3M(R)}\right)^{1/3},
    \label{eq:tidalradius}
\end{equation}

\noindent where $R$ is the galactocentric distance, $m$ the progenitor mass, and
$M(R)$ the enclosed Galactic mass. This expression is the Jacobi radius
for a progenitor on a circular orbit around a point mass. For an extended
mass distribution the factor $3$ becomes $3 - d\ln M/d\ln R$, which equals
$2$ for an isothermal halo with $M \propto R$. Variations in $r_t$ across the
halo provide direct insight into the properties of its mass distribution. This form
follows from balancing the progenitor's self-gravity against the
differential (tidal) component of the Galaxy's gravitational force
across the progenitor's extent; stars beyond $r_t$ are more strongly
pulled by the Galaxy's tidal field than they are bound to the
progenitor, and so become unbound. The associated
orbital energy dispersion of stream stars can be written as
\end{sloppypar}
\begin{equation}
    \Delta E \sim \frac{\partial\Phi}{\partial r}\cdot\Delta r,
    \label{eq:energydispersion}
\end{equation}

\noindent linking local gradients in the potential to observable
fragmentation and misalignment along the stream. Smooth, symmetric halos
therefore produce streams that remain narrow and continuous, while more
complex potentials generate distortions, gaps, and misalignments
\citep{koposov2010}. Encounters with dark matter subhalos leave localized
signatures such as spurs or gaps \citep{bonaca2019, pricewhelanbonaca2018},
making streams one of the most powerful tools for testing cosmological
predictions, as well as the microphysical assumptions
underlying different dark matter candidates, against the reality of the
Milky Way's dark matter halo.

\section{\hspace{0cm}Observational Signatures of the Galactic Halo in Stellar\newline Streams}
\label{sec:Signatures}

\subsection{Coherence and Continuity}

Stellar streams are expected to remain
narrow, coherent, and continuous along their orbits in smooth, symmetric halo models 
(see Equations~\ref{eq:cbe}--\ref{eq:energydispersion}). Every star that escapes the progenitor moves on a similar orbit in the same static potential, tracing a single, well-defined track on the sky with a spatial extent governed by the small spread in energy and angular momentum upon escape. However, observations show variations in stream thickness and
density with several known streams appearing fragmented,
kinked, or misaligned with the smooth orbit a simple model would predict
\citep{koposov2010, grillmairdionatos2006}. A cold stream has few ways of disturbing its own path through internal
processes. Warps and fragmentation therefore often point to features of
the Galactic potential that simple models do not capture. Internal effects
can still matter, because stars escaping the progenitor form epicyclic
overdensities near it, which is why the progenitor must be modeled as well
(Section~\ref{sec:Sensitivity}). Identifying the true origin of these irregularities is the primary goal
of the remainder of this section and the case studies that follow.
Possible causes include global triaxiality, an evolving potential, bar
resonance, a close encounter with a compact perturber, and epicyclic
substructure from the progenitor itself.

\subsection{Gaps and Spurs}

\begin{figure}[htbp]
\centering
\includegraphics[width=0.78\textwidth]{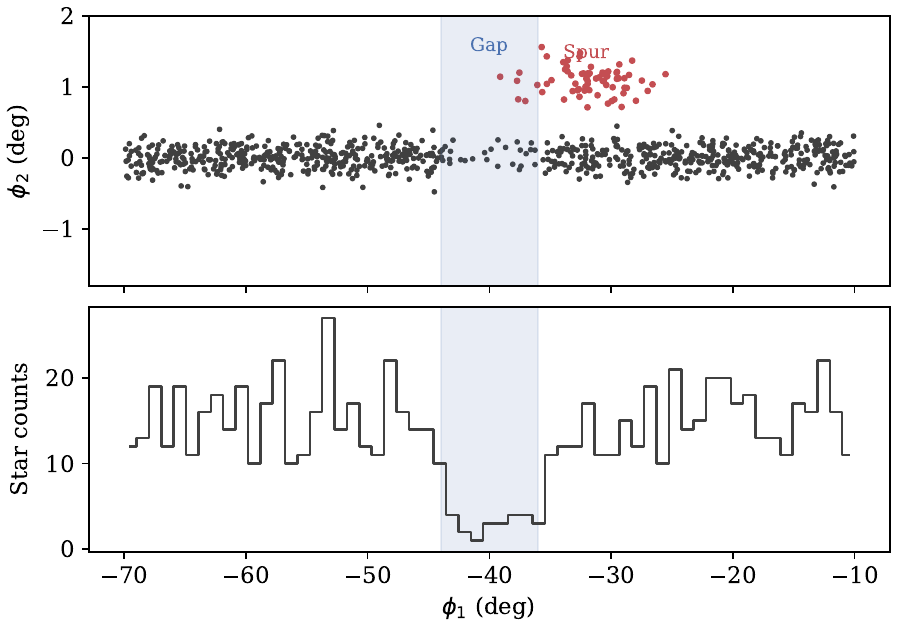}
\caption{Schematic illustration of a subhalo-induced gap and spur in a
stellar stream. \textit{Top:} stream stars in stream-aligned coordinates
$(\phi_1,\phi_2)$; a localized encounter depletes stars in a narrow range
of $\phi_1$ (gap, shaded) and deposits an offset clump of stars downstream
(spur). \textit{Bottom:} the corresponding dip in linear stellar density
across the gap region. This figure is a schematic illustration of the
mechanism and is not derived from any specific published dataset.}
\label{fig:gapspur}
\end{figure}

When observing stellar streams, morphological features such as gaps and spurs provide some of the clearest
signatures of dark matter substructure. A gap is a region where stars are
missing along the stream, while a spur is a branch of stars offset from
the main path of the stream (Figure~\ref{fig:gapspur}). The GD-1 stream is the
prominent example: \citet{bonaca2019} identified both a spur and a gap
consistent with an encounter between the stream and a dark matter subhalo
in the mass range $10^{6}$--$10^{8}\,M_\odot$, a finding backed up later by dedicated spectroscopic observations \citep{bonaca2020}. The
velocity perturbation experienced by stream stars during such an impulsive encounter
can be approximated by

\begin{equation}
    \Delta v \approx \frac{2GM}{bV},
    \label{eq:velocitykick}
\end{equation}

\noindent where $M$ is the mass of the perturber, $b$ the impact
parameter, and $V$ the relative encounter velocity. This expression
relies on the impulse approximation, where the star's position is treated as fixed during the encounter, allowing the velocity change to be expressed as the time integral of the perpendicular gravitational force
exerted by the passing perturber. The resulting spatial
offsets and density variations appear as gaps and spurs.
\citet{erkalbelokurov2015} established a framework for how gap growth depends on subhalo mass and encounter time. Their work showed that gap depth and width can be mapped back to the properties of the perturbing subhalo, precisely the kind of
subhalo-abundance measurement needed to distinguish
between the dark matter candidates discussed in
Section~\ref{sec:dmproblem}. However, gaps do not always originate from dark matter subhalos; \citet{pearsonpricewhelanjohnston2017} showed that resonant interactions with the Galactic bar can produce gaps and length asymmetries
in the Palomar 5 stream, demonstrating that baryonic structures must be
modeled carefully before subhalo interpretations are adopted. Despite this, these features still provide direct observational evidence that stellar streams are sensitive to small-scale structure in the halo. As a result, they can detect low-mass dark matter subhalos in our own
Galaxy, a regime that is difficult to reach with other current techniques.

\subsection{Wiggles and Bending}
\begin{sloppypar}
    
In addition to localized disruptions such as gaps and spurs, streams
can show broader, more gradual changes in shape. The stream may shift
slightly from side to side, its width or distance from the progenitor
may change slowly along its length, or the entire stream may deviate
systematically from the expected reference orbit. While gaps reflect a
discrete, localized encounter, these broad-scale variations are
directly sensitive to halo properties that extend continuously along
the entire stream. A triaxial halo, for
instance, imprints a characteristic large-scale wave in a stream's track
because the direction and strength of the restoring force varies smoothly
with orbital phase around a non-spherical potential. A resonance with the
rotating Galactic bar can similarly produce a slow, coherent bending
concentrated in the inner Galaxy \citep{pearsonpricewhelanjohnston2017}.
Time-dependent perturbations from massive, still-infalling satellites such
as the Large Magellanic Cloud act on a comparable scale, but change with
time rather than being fixed in the Galactic frame, so their effect can in
principle be distinguished from a static triaxial halo by comparing
streams on different orbits or at different Galactocentric distances
\citep{erkal2019lmc}. Wiggles and bending encode the geometry and
time-dependence of the potential over an extended volume, rather than
the presence of a single perturber. They therefore complement gaps and
spurs as a distinct and largely independent constraint on the overall
shape of the dark matter halo.
\end{sloppypar}

\subsection{Case Studies}
\label{sec:Case}
\begin{sloppypar}

We consider five streams that together span a broad range of progenitor
types, orbital geometries, and halo regimes, deliberately drawing on
several rather than one so that the pattern across systems, and not any
single stream's idiosyncrasies, drives our conclusions about the halo.

\begin{figure}[htbp]
\centering
\includegraphics[width=0.85\textwidth]{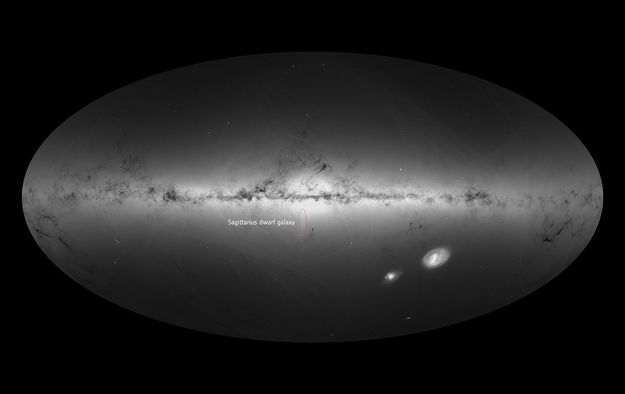}
\caption{All-sky map of the density of stars observed by \textit{Gaia}
between July 2014 and May 2016. The Sagittarius dwarf galaxy is visible as an elongated overdensity below the Galactic center. \emph{Credit: ESA/Gaia/DPAC. Licensed under CC BY-SA 3.0 IGO.}}
\label{fig:sgr-gaia}
\end{figure}

\textbf{Sagittarius Stream:} Discovered as a disrupting dwarf spheroidal
galaxy by \citet{ibata1994}, the Sagittarius stream was later mapped
across the sky using Two Micron All Sky Survey photometry by
\citet{majewski2003} and Sloan Digital Sky Survey photometry by
\citet{belokurov2006streams}. The Sagittarius dwarf galaxy is also visible in the Gaia all-sky map
(Figure~\ref{fig:sgr-gaia}). Extending more
than $360^{\circ}$ around the Galaxy on a polar, eccentric orbit, the stream has
been used extensively to test halo shape
\citep{johnstonlawmajewski2005, lawjohnstonmajewski2005}, and it remains
the single most intensively modeled halo tracer in the Galaxy: as
discussed in detail in Section~\ref{sec:Halo Models}, successive triaxial fits to its debris have favored progressively more complex
halo shapes as additional physics, particularly the time-dependent
influence of the LMC, was incorporated~\citep{lawmajewski2010, degwidrow2012, veraciroaelmi2013, vasilievetal2021}.

\textbf{GD-1 Stream:} First detected by \citet{grillmairdionatos2006} as
a narrow, $\sim 63^{\circ}$ long tidal stream from a disrupted globular cluster, GD-1 was later mapped in
six-dimensional phase space by \citet{koposov2010}, who found it lies at
a heliocentric distance of only $7$--$10\,\mathrm{kpc}$ on a retrograde
orbit with perigalacticon $\sim14\,\mathrm{kpc}$ and apogalacticon
$26$--$29\,\mathrm{kpc}$, and it was further refined with \textit{Gaia}
astrometry by \citet{pricewhelanbonaca2018}, who extended the mapped
stream by a further $\sim 20^{\circ}$ to its now commonly cited total
length of $\sim 80^{\circ}$. Because it lacks a surviving
progenitor and stays well clear of the Galactic bar and disk for most of
its orbit, GD-1 is an unusually clean laboratory for studying small-scale perturbations. Using Gaia DR2 astrometry together with spectroscopic measurements, \citet{malhanibata2019halo} constrained the Galactic potential with GD-1, including the flattening of the Milky Way dark matter halo. \citet{malhanetal2019cocoon} additionally identified a broad, diffuse stellar ``cocoon surrounding its narrow stream track, consistent with models in which the progenitor formed in a dark matter subhalo before accretion.  Its spur and gap features \citep{bonaca2019, bonaca2020} highlight its
sensitivity to encounters with compact perturbers, while later studies have also explored the possibility that the Sagittarius dwarf galaxy could have contributed to the observed spur \citep{deboer2020gd1, bonaca2020}. Modeling of its stellar-density perturbations has been used to constrain the abundance of dark subhalos and the particle nature of dark matter \citep{banik2021}, providing a connection between stellar-stream observations and the distinction between warm and cold dark matter introduced in Section~\ref{sec:dmproblem}.

\begin{figure}[htbp]
\centering
\includegraphics[width=0.85\textwidth]{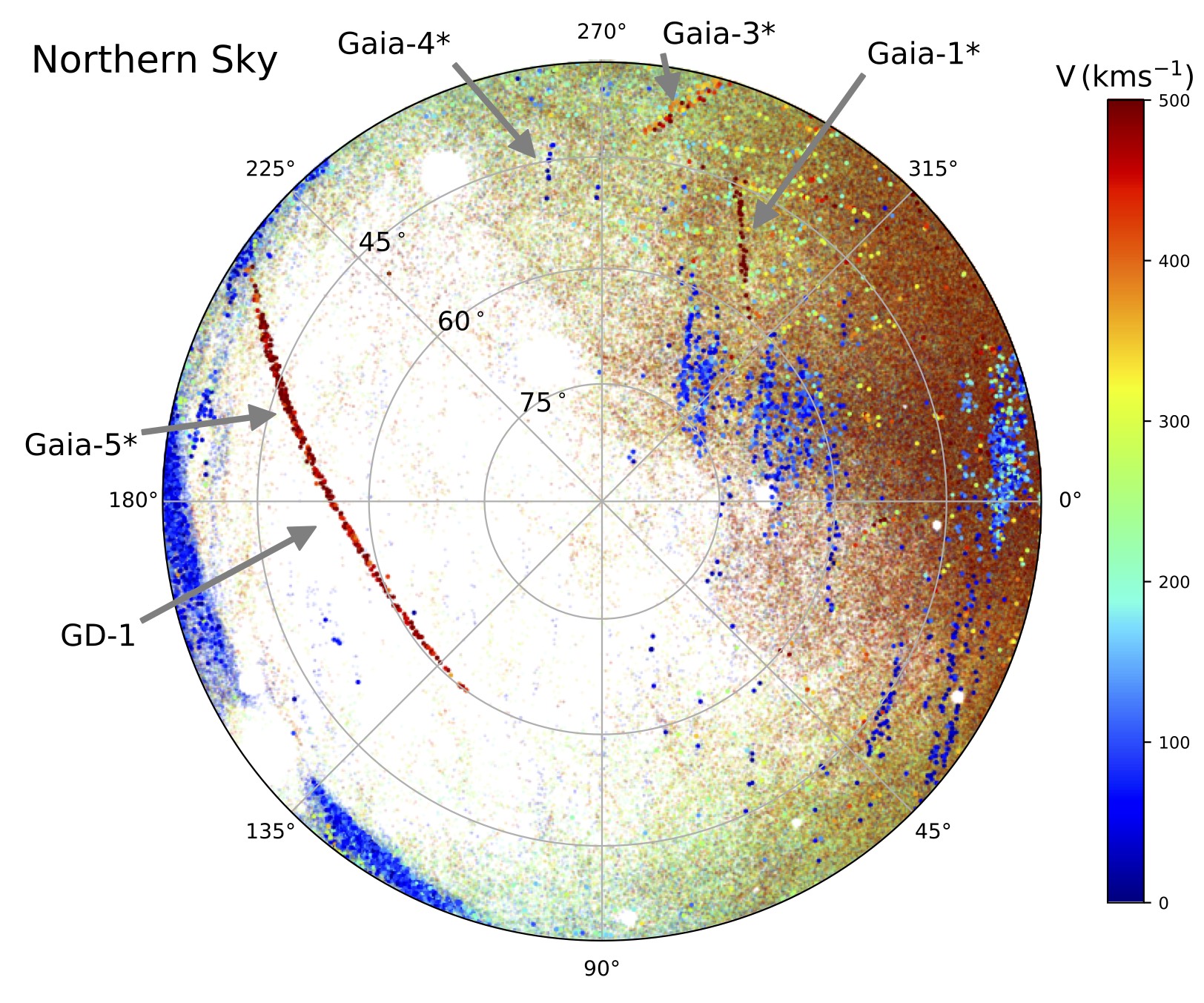}
\caption{STREAMFINDER map of stellar-stream structures in the
northern Galactic hemisphere, identified from \textit{Gaia} DR2 data
using the algorithm of \citet{malhanibata2018}. The GD-1 stream is
labeled, along with several streams reported as new discoveries at the
time (asterisked). Point color encodes stellar flow speed. \emph{Credit:
ESA/Gaia/DPAC; K. Malhan, R. Ibata \& N. Martin, Observatoire
astronomique de Strasbourg} \citep[cf.][]{ibata2021streamfinder}.}
\label{fig:gd1-gaia}
\end{figure}

\textbf{Palomar 5 Stream:} Discovered by \citet{odenkirchen2001}, the Palomar 5 stream arises from a globular cluster on a moderately eccentric, prograde orbit with
perigalacticon $\sim7\,\mathrm{kpc}$ and apogalacticon
$\sim19\,\mathrm{kpc}$. The tails
were first traced over about $10^\circ$ by \citet{odenkirchen2003} and
later to $\sim22^{\circ}$ by \citet{grillmairdionatos2006b}. Unlike GD-1, its two tidal tails are
markedly asymmetric, the trailing tail extends roughly $18.5^{\circ}$
while the leading tail reaches only about $3.5^{\circ}$. The stream's
morphology and clumpy density variations provide constraints on both the
smooth halo component \citep{bovybahmanyarfritzkallivayalil2016} and
possible perturbations driven by the bar or a subhalo 
\citep{pearsonpricewhelanjohnston2017, kupper2015}, with the leading and trailing tail asymmetry itself often attributed to resonant interaction with
the rotating Galactic bar rather than to a discrete dark matter
encounter.

\textbf{Orphan--Chenab Stream:} Independently discovered by
\citet{grillmair2006orphan} and \citet{belokurov2007orphan}, and later
given a well-defined orbit by \citet{newberg2010}, who found a
comparatively eccentric orbit ($e \approx 0.7$) with perigalacticon
$\sim16\,\mathrm{kpc}$ and apogalacticon $\sim90\,\mathrm{kpc}$, carrying
it much farther into the outer halo than GD-1 or Palomar 5. \citet{koposov2019orphan} mapped the Orphan-Chenab stream across the full
sky. They showed that its southern part had already been found in the Dark
Energy Survey and named Chenab \citep{shipp2018}. \citet{shippdesipomotions2019}
then measured the proper motion of Chenab independently. The stream's
kinematic misalignment with its orbital track gave the first clear
dynamical detection of the Large Magellanic Cloud's gravitational
influence on the outer halo \citep{erkal2019lmc}. A similar signal was
later found in the Sagittarius stream itself \citep{vasilievetal2021}, as
discussed in Section~\ref{sec:challenges}.

\textbf{Cetus Polar Stream:} \citet{newberg2009cetus} discovered this stream
kinematically in Sloan Extension for Galactic Understanding and
Exploration (SEGUE) velocities. It runs along $l \approx 143^\circ$ in the
South Galactic Cap on a nearly polar orbit. While initially treated as sparsely characterized, recent spectroscopic and distance mapping by \citet{thomas2022cetus} has traced the system (which they propose to rename the Cetus-Palca stream) across an extent of $\sim 100^\circ$ with heliocentric distances spanning $25\text{--}40\text{ kpc}$ and a stellar mass pointing to a disrupted dwarf galaxy remnant. Its extended polar orbit makes it a powerful probe of the outer-halo potential.

\begin{table}[htbp]
\begin{adjustwidth}{-1.8cm}{0cm}
\centering
\caption{Case studies of Milky Way stellar streams discussed in this review.}
\label{tab:streams}

\footnotesize
\begin{tabular}{p{2.3cm} p{2.3cm} p{1.3cm} p{2.6cm} p{1.0cm} >{\centering\arraybackslash}p{1.8cm} >{\raggedright\arraybackslash}p{3.2cm} >{\raggedright\arraybackslash}p{2.8cm}}
\toprule
Stream & Progenitor & Extent & Orbit type & $d_\odot$ (kpc) & $l$ & Key features & References \\
\midrule
Palomar~5 & Globular\newline cluster & $\sim22^\circ$ & Prograde, moderately\newline eccentric & $\sim23$ & $\sim0.9^\circ$ & Wide tails; bar-driven gap asymmetry & \citet{odenkirchen2001, grillmairdionatos2006b, pearsonpricewhelanjohnston2017} \\
[0.5em]
Sagittarius & Dwarf galaxy & $>360^\circ$ & Polar, eccentric & $15$--$100$ & $0^\circ$--$360^\circ$ & Multiple wraps; sensitive to Galactic halo shape & \citet{ibata1994, majewski2003, lawmajewski2010, hernitschek2017} \\
[0.5em]
GD-1 & Globular\newline cluster & $\sim80^\circ$ & Retrograde, thin & $7$--$10$ & $120^\circ$--$220^\circ$ & Narrow track; prominent gaps and spur (possibly due to perturbations) & \citet{grillmairdionatos2006, koposov2010, bonaca2019} \\
[0.5em]
Orphan–Chenab & unknown; Likely dwarf\newline galaxy remnant & $\sim210^\circ$ & Highly inclined & $15$--$60$ & $170^\circ$--$250^{\circ\star}$ & LMC-induced track misalignment & \citet{grillmair2006orphan, belokurov2007orphan, newberg2010, erkal2019lmc, koposov2019orphan} \\
[0.5em]
Cetus \newline Polar & Dwarf galaxy remnant & $\sim100^\circ$ & Near Polar &  $25$--$40$ & $\sim143^\circ$ & Narrow polar stream; probe of the outer-halo potential & \citet{newberg2009cetus, thomas2022cetus} \\
\bottomrule
\end{tabular}
\end{adjustwidth}
\end{table}

\vspace{-0.9em}
{\footnotesize
\noindent\hangindent=0.5em\hangafter=1
$^{\star}$The quoted longitude range corresponds to the portion of the Orphan stream covered by its first observations. Later studies have traced the stream across a much larger part of the sky.\par}
\vspace{0.5em}
The population-level power of this approach extends well beyond these five case studies. For instance, analyzing large samples of stream orbits simultaneously, such as the 23-stream clustering study by \citet{bonaca2021orbital}, reveals that nearly all examined streams have an ex situ origin:  most trace back to accreted dwarf galaxies, while eight have likely globular cluster progenitors. Some streams that look spatially
distinct also appear to share a common progenitor. This is exactly the
kind of population-scale inference that motivates considering multiple
streams jointly rather than in isolation.

These five systems cover a broad range of progenitor types, orbits, and
orbital orientations. The progenitors include both dwarf galaxies and
globular clusters. The orbits range from Palomar 5's moderately eccentric
path to the Orphan-Chenab stream's orbit extending far out into the halo,
and the orbital planes include prograde, retrograde, and polar cases.
\begin{sloppypar}
Stellar streams are not simple relics of disrupted systems. They serve as active probes of the Galactic potential, directly tracing the dark matter that dominates it.
For a broader view, see the review of \citet{newbergcarlin2016} and the
nearly hundred stream compilation of \citet{mateu2023}.
\end{sloppypar}
\noindent \textit{Note} that the heliocentric distance ($d_\odot$) and Galactic longitude
($l$) columns describe the observed extent of each stream. This generally
differs from the Galactocentric perigalacticon and apogalacticon of its
orbit, which are given in the text above where available.
\end{sloppypar}

\section{Methodological Approaches}
\label{sec:Methodology}

Stellar streams are used to study the dark matter halo of the Galaxy, and many different methods have been developed for this. These methods connect theoretical models with observations. They range from simple geometric orbit fits to full statistical inference on large stellar catalogs. These approaches did not emerge all at once. Each responded to a specific limitation of the one before it. Over time, these limits pushed researchers from simple orbit fits toward the detailed models used today.

\subsection{Orbit Fitting and Analytical Models}

The earliest methods fitted the track of a stream with an orbit in an assumed Galactic potential \citep{johnstonzhaospergelhernquist1999}. The idea is simple. A single test particle is integrated along an orbit in a trial potential, and this orbit is compared directly with the observed track of the stream on the sky. When distances and velocities along the stream are available, they are compared too. The parameters of the potential are then adjusted until the fit is acceptable.

This approach was used widely on the Sagittarius stream, with data from the Two Micron All Sky Survey \citep{lawjohnstonmajewski2005}. Earlier studies were more restricted in scope, and their axisymmetric fits gave conflicting results \citep{helmi2004}. \citet{lawmajewski2010} found that the spatial and kinematic structure of the stream could be reproduced by a triaxial halo potential. This result helped to resolve the earlier conflict between the axisymmetric fits.

Orbit fitting provides a useful first approximation and remains computationally inexpensive. However, it generally neglects the effects of progenitor disruption and phase mixing. It treats the stream as if it were a single orbit rather than a group of many slightly different orbits. Real streams are more complex than any single orbit, and a better fit would not change that. A real stream has finite width and a range of internal energies, so a single-orbit fit implicitly averages over structure that may itself carry information about the potential. It also cannot account for features such as gaps or spurs that arise from the differential evolution of individual stream stars. This gap between what single-orbit fitting can explain and what real streams actually look like motivated the shift toward full N-body and test-particle modeling described next.

\subsection{N-body and Test-Particle Simulations}
\begin{sloppypar}
\setlength{\parskip}{0.5em}
Numerical simulations offer a more realistic representation of stream
formation than a single fitted orbit. Full N-body models track the
gradual tidal disruption of progenitors directly
\citep{johnstonhernquistbolte1996}. The progenitor is followed as it loses mass over many orbits, with its own self-gravity included. Stars spread out along nearby orbits, and the model captures this spread, including features such as gaps or wiggles that a single-orbit fit cannot produce.
However, a full N-body simulation of a progenitor and its surrounding stream is computationally expensive. This is especially true when many trial potentials must be explored to fit an observed stream. For this reason, more efficient approximations have been developed. Streakline and
``particle-spray'' methods release many massless test
particles near the progenitor's tidal radius. New particles are released at each point along the progenitor's orbit. This gives a simple model of tidal stripping, without modeling the
progenitor's internal dynamics in detail
\citep{vargheseibatalewis2011}. These particles are released with
a small but systematic offset in orbital energy relative to the
progenitor. As the integration proceeds, this offset makes the particles separate naturally into leading and trailing tails. The result reproduces the basic morphology of a real
stream at a small fraction of the computational cost of a full N-body
run. The approach can also be combined with efficient action-angle methods, which are available in
open-source dynamics packages \citep{bovy2015galpy,
price-whelangala2020}. Together, these approaches enable the systematic exploration of stream tracks across multiple candidate potentials, which is essential for the statistical inference methods described below.
\end{sloppypar}

\subsection{Statistical Inference Frameworks}

\begin{sloppypar}
Newer work no longer relies on one best-fit model. Instead, it uses a
likelihood-based measure of how well the stream properties predicted by a
model match the observed ones. \citet{sandersonhelmihogg2015} introduced an action-space
clustering approach to infer the Galactic potential directly from
phase-mixed tidal debris. The idea is that stars from a common progenitor
should cluster tightly in action space only when the assumed potential is
close to the true one. \citet{bovybahmanyarfritzkallivayalil2016} combined
the phase-space tracks of the Pal 5 and GD-1 streams in a Bayesian
dynamical model. This constrained the shape and mass of the inner Galactic
halo at the same time. The axis ratio they found differs somewhat from
simple $\Lambda$CDM expectations. \citet{bonacahogg2018} used an information-theoretic approach to quantify
how much a single stream can realistically tell us about the potential.
The same analysis shows how many streams are needed, and how well each
must be observed, to break the degeneracies discussed in
Section~\ref{sec:challenges}. \citet{bonaca2021orbital} then applied orbital clustering in energy and
angular momentum space to 23 Galactic streams at once. Most of the streams
could plausibly be linked to a specific disrupted dwarf galaxy that
brought them into the Milky Way, and for eight of them a likely globular
cluster progenitor was also identified. This conclusion could not have
been reached from a single stream. It also shows why this review treats
streams as a population rather than as individual case studies. Such
clustering can also be used to constrain the Galactic potential, so the
same population-level approach helps with inference.
Likelihood-based frameworks are what make simultaneous analysis of many
streams possible. Each stream's constraint on the potential can be written
as a probability distribution over shared parameters. Constraints from
different streams can then be combined statistically, instead of being
reconciled manually. This matters more as the number of well-measured
streams grows from a few to many. These methods work especially well with
large, homogeneous datasets. They allow halo shape, substructure and time
dependence to be constrained together, rather than treated as separate
fitting problems.
\end{sloppypar}
\subsection{Role of Astrometric and Photometric Surveys}
\begin{sloppypar}
\setlength{\parskip}{0.5em}
Advances in observational catalogs have transformed the study of stellar
streams from a search for a small number of exceptionally clear cases into a
systematic, population-based approach, and much of this shift has been
driven as much by new detection algorithms as by deeper photometric and astrometric surveys. The
\textit{Gaia} mission's second and third data releases
\citep{gaiadr2brown2018, gaiadr3} provided proper motions and parallaxes
for well over a billion stars, but leveraging this volume of data
required moving beyond the visual search methods and matched-filter techniques used to
find earlier streams such as GD-1 and the Orphan-Chenab stream. The
STREAMFINDER algorithm \citep{malhanibata2018} addressed this by
searching simultaneously in position, kinematic, and photometric space
for stars consistent with lying along a common orbit, without requiring prior selection of candidate streams (Figure~\ref{fig:gd1-gaia}). Applied to \textit{Gaia} data, it has since been used to produce large,
homogeneous catalogs of stream members across the whole sky
\citep{ibata2021streamfinder}, including the catalog underlying several
of the case studies discussed in Section~\ref{sec:Case}. 

In practice, isolating a stream's members from the far larger foreground
and background stellar population combines several of \textit{Gaia}'s
data products in sequence. Photometry in the $G$, $G_{BP}$, and $G_{RP}$
bands is first used to select stars consistent with a single old,
metal-poor isochrone at a plausible distance, since stars from a globular cluster share a common age and metallicity (streams from dwarf galaxies span a wider range, which calls for a broader selection); this photometric selection is then combined
with \textit{Gaia}'s proper motions, which for a cold, kinematically
coherent stream cluster tightly in a small region of proper-motion space
distinct from the broader, dynamically hotter field population. Parallaxes
are useful mainly for nearby streams, given that \textit{Gaia}'s parallax
precision degrades rapidly beyond a few kiloparsecs, while radial velocities from the Radial Velocity Spectrometer (RVS) are
available only down to $G_{\rm RVS} \approx 12\,$mag in DR2 and
$G_{\rm RVS} \approx 14\,$mag in DR3 \citep{katz2023}. This is roughly six magnitudes shallower than \textit{Gaia}'s astrometric and photometric limit. Full 6D phase-space coverage
is therefore restricted to comparatively bright and much smaller samples, even for well-studied streams such as GD-1 \citep{koposov2010}. Algorithms such as STREAMFINDER
\citep{malhanibata2018} automate this combination of photometric and kinematic selection. They apply it across the whole sky, so candidate streams no longer require manual selection.

Complementary photometric surveys, including Pan-STARRS1 \citep{panstarrs2016} and the Dark Energy
Survey \citep{shipp2018, shippdesipomotions2019}, have added many streams to the known sample beyond the regions and magnitude limits that Gaia data can probe, particularly in the southern sky. The Vera C. Rubin Observatory's Legacy Survey of Space and Time
\citep{lsst2019} began in 2026. It is expected to extend this further,
widening the dataset and reaching substantially fainter magnitudes than current surveys. Analysis pipelines built on
open-source scientific software \citep{price-whelan2018astropy} have been necessary for combining these
heterogeneous datasets into consistent dynamical models. This is because
no single survey currently provides position, distance, proper motion, and
radial velocity with comparable precision across the whole sky. These
observational and algorithmic advances give the data against which the
theoretical models and statistical frameworks described above are tested.
They are also the main reason the field has moved, within roughly a
decade, from studying a few isolated streams to characterizing streams as
a statistical population.
\end{sloppypar}
\begin{sloppypar}
\section{Challenges and Open Questions}
\label{sec:challenges}

\subsection{Degeneracies in Halo Modeling}

A major difficulty in interpreting stellar streams is the degeneracy
between different halo properties. \citet{bonacahogg2018} used an
information-theoretic framework to show quantitatively that a single cold
stream constrains well only a limited combination of the potential parameters.
This is typically some combination of enclosed mass and local shape along
the orbit the stream happens to occupy, and not the full three-dimensional
structure of the halo. It is an instance of a more general problem
familiar from other areas of galactic dynamics. A well-known example is the
mass-anisotropy degeneracy in spherical Jeans modeling, where different
combinations of enclosed mass and orbital anisotropy can reproduce the
same observed velocity dispersion profile. A single dynamical measurement
rarely determines more than one or two effective combinations of the
underlying physical parameters. Breaking the degeneracy between halo
shape, mass, and substructure therefore generally requires combining
multiple independent streams. These streams should probe different regions
of the Galaxy, at different distances and orbital inclinations, and ideally
have complementary sensitivities \citep{bonacahogg2018, sandersonhelmihogg2015}. This is
why this review considers several streams together, instead of treating
any one stream as conclusive on its own.

\subsection{Limited Observational Coverage}

Prominent streams such as Sagittarius, GD-1, Palomar 5, and the Orphan
stream have been the subject of many studies. Most other streams discovered in the
Milky Way are much more poorly characterized than these examples. The
systematic compilation of \citet{mateu2023} lists nearly a hundred distinct
stellar streams around the Galaxy. Only a small fraction of these have the
full six-dimensional phase-space coverage (sky position, distance, proper
motion, and radial velocity) needed for the detailed dynamical modeling
described in Section~\ref{sec:Methodology}. Many of the remaining streams
are faint, sparsely populated, or lie in regions obscured by dust and
crowding in the Galactic disk. This limits both the number of member stars
that can be confidently identified and the precision with which their
distances, proper motions, and radial velocities can be measured. Streams
confined to the southern sky, or to regions not yet covered by deep
photometric surveys, are correspondingly less well constrained than their
northern, well-surveyed counterparts. This observational incompleteness is
not simply a matter of having fewer data points for a given stream. More
fundamentally, it limits our ability to draw robust conclusions about the global properties of the halo. The streams that are bright and clean enough for detailed
dynamical modeling may not be a representative sample of the full
population in orbit, progenitor type, or Galactocentric distance.

\subsection{Sensitivity to Progenitor Properties}
\label{sec:Sensitivity}
Stream morphology depends not only on the Galactic potential but also on
the properties of the progenitor system. The progenitor must therefore be
modeled alongside the potential, instead of being treated as a fixed,
known input. \citet{kupper2015} demonstrated that this holds even for a
well-studied system like Palomar 5. Separating the progenitor's internal
dynamics (its mass, concentration, and internal velocity dispersion at the
time of disruption) from the imprint of the external potential requires
careful, high-precision modeling of the progenitor and the potential at
the same time. Differences in progenitor mass, orbit, and internal
structure can therefore make it harder to isolate halo signatures from
progenitor-driven effects. This is especially true for streams whose
progenitor has already been fully disrupted, because it can no longer be
observed directly to constrain its original properties. The problem is most acute for dynamically old, fully phase-mixed streams,
where no bound remnant is left to constrain the model. It matters least
for dynamically young, only partially disrupted streams such as
Palomar 5, where the progenitor cluster can still be observed directly
alongside its tidal tails.

\subsection{Time-Dependent Dynamics}

The Milky Way halo is not static. Treating it as static can lead to
systematically incorrect conclusions. \citet{erkal2019lmc} showed that the
observed misalignment between the Orphan-Chenab stream's track and its
proper motion cannot be explained by any realistic static Galactic
potential. The misalignment is readily reproduced once the gravitational
influence of an infalling Large Magellanic Cloud of mass
$\sim 1.4\times10^{11}\,M_\odot$ is included. Similar reflex-motion misalignments have since been found in the Sagittarius stream itself \citep{vasilievetal2021}. This indicates that the effect is not a peculiarity
of one system. It is a general signature of the LMC's ongoing infall,
imprinted across multiple streams on different orbits. Together,
these results demonstrate that the LMC has a measurable dynamical effect
on halo streams extending well beyond its immediate neighborhood. Accounting for such time-dependent perturbations is essential for
accurate modeling of the outer halo, since a static-potential fit to a
stream affected by the LMC's passage risks misattributing a transient
perturbation to permanent halo triaxiality or substructure. Baryonic
structures such as the rotating Galactic bar introduce a comparable
complication in the inner halo \citep{pearsonpricewhelanjohnston2017},
underscoring that both external (LMC) and internal (bar) time-dependence
must be modeled before subhalo or triaxiality interpretations of a given
feature can be considered secure.
\end{sloppypar}

\subsection{Small-Scale Structure and the Nature of Dark Matter}
\begin{sloppypar}
Halo substructure matters because of a set of long-standing small-scale tensions in $\Lambda$\text{CDM}. Cosmological CDM-only simulations predict
far more low-mass subhalos around a Milky Way-sized galaxy than the
number of observed dwarf satellite galaxies, the so-called missing
satellites problem \citep{klypin1999, moore1999}. The most massive of
the predicted subhalos, moreover, appear too dense to host any of the
Milky Way's known bright satellites without overproducing their observed
stellar velocity dispersions, a related discrepancy known as the
too-big-to-fail problem \citep{boylankolchin2011}. A third, longer-standing puzzle is the core-cusp problem. CDM simulations
predict halos with steeply rising (cuspy) central density profiles.
Rotation curves of many dwarf and low-surface-brightness galaxies instead
favor a flatter core with a roughly constant density \citep{deblok2010}.
The internal structure of accreted globular-cluster streams offers a complementary probe of these density profiles: \citet{malhanetal2022profiles} compared Gaia EDR3 proper motions with simulations of clusters originating in cuspy and cored dwarf-galaxy subhalos, finding that their tangential velocity dispersions can constrain the density profiles of their former hosts. Baryonic feedback can resolve some of these discrepancies within CDM
itself. Each issue remains actively debated, however, and ultimately
reflects the abundance and internal structure of low-mass halos and
subhalos. These are the objects that gaps and spurs in stellar streams are
sensitive to. Stream-based subhalo counts therefore bear on these debates.
They give an observational probe of the same small-scale regime, using a
method that is largely independent of satellite counting or rotation-curve
fitting. Perhaps the most important open question, and the one that ties
this review most directly back to the dark matter problem introduced in
Section~\ref{sec:dmproblem}, is how stream signatures connect to the fundamental nature of dark matter. Gaps and spurs may reflect encounters
with subhalos, but the predicted abundance and density profile of these
subhalos depend on the underlying dark matter model \citep{banik2018}:
cold dark matter predicts a steep, largely unsuppressed mass function of
subhalos down to very low masses, while warm or self-interacting dark
matter suppress or reshape that population below a characteristic scale
set by the particle's free-streaming length or interaction cross-section.
Using the density profile of the GD-1 stream, \citet{banik2021} placed
constraints on the free-streaming (half-mode) mass of the dark matter
power spectrum, demonstrating how stream substructure can discriminate between cold, warm
\citep{bodeostrikerturok2001}, and self-interacting
\citep{spergelsteinhardt2000} dark matter scenarios using a single
Galactic stream. The statistical power of this approach should improve
substantially once it is applied jointly to the much larger stream
sample now being assembled \citep{mateu2023, ibata2021streamfinder},
rather than to one or two well-studied cases. It remains an open challenge to determine whether current and future
stream samples can robustly distinguish between these scenarios. This
requires analyzing the streams collectively, as a statistical population,
instead of one stream at a time. The goal is confidence comparable to that
of laboratory direct-detection limits or cosmological structure-formation
constraints.
\end{sloppypar}

\section{Comparison of Halo Models}
\label{sec:Halo Models}

\subsection{Idealized Models}
\enlargethispage{0.5\baselineskip}
Early studies of the Milky Way's dark matter halo often relied on
simplified representations, adopting spherical or axisymmetric
density profiles, such as the singular isothermal sphere or the NFW
profile 
\citep{nfw1996, nfw1997}. Such models are mathematically
tractable. They need only a minimal number of free parameters (typically a
scale radius along with a characteristic density or mass). They also admit
closed-form expressions for the potential and the circular velocity curve, and their orbits are simple enough to treat analytically even where a
fully closed-form solution is not available. This simplicity made them the
usual starting point for the earliest dynamical studies of the Galaxy.
They remain useful today as a well-understood baseline against which more
complex models can be compared. The rotation curve in 
Figure~\ref{fig:rotcurve}, for
instance, is itself built from a simplified combination (a spherical NFW
halo plus an axisymmetric disk and bulge). These assumptions limit the
complexity of the resulting dynamics. A strictly spherical or axisymmetric,
static halo does not naturally produce the gaps, spurs, or large-scale
wiggles described in Section~\ref{sec:Signatures}.
In the spherical case, orbital planes are
fixed, although apsidal precession can occur within those planes. More
generally, the resulting orbital evolution remains smooth, regular, and
fully predictable.

\subsection{Realistic Galactic Potentials}
\begin{sloppypar}
Cosmological simulations in the $\Lambda$CDM paradigm instead predict halos that
are triaxial and rich in substructure, reflecting their hierarchical
assembly history. \citet{jingsuto2002} used high-resolution
N-body simulations to show that halo density profiles are better described by a
triaxial generalization of the NFW profile. The axis ratios of such halos vary systematically with halo mass and with distance
from the halo center. Halos are typically more elongated in the inner regions and
closer to spherical in the outer regions. The Sagittarius stream is a useful, and somewhat cautionary case study of what
happens when this theoretical expectation meets real data.
\citet{lawmajewski2010} found that fitting the stream's full spatial and
kinematic structure over $20$-$60\,\mathrm{kpc}$ required a triaxial
potential with minor-to-major and intermediate-to-major axis ratios of
$(c/a)_\Phi = 0.72$ and $(b/a)_\Phi = 0.99$ respectively. This represents a strongly
flattened halo. Its best-fit orientation placed the disk nearly
perpendicular to the halo's intermediate axis, a configuration
later shown to be dynamically unstable over the disk's lifetime
\citep{veraciroaelmi2013}. Independent MCMC modeling by
\citet{degwidrow2012} found comparably strong triaxiality, with axis
ratios of $3.3\pm0.7$ and $2.7\pm0.4$. They noted, however, that such a strongly
aspherical halo is difficult to reconcile with standard $\Lambda$CDM
structure-formation predictions at the relevant halo mass. This tension
prompted alternative models. \citet{veraciroaelmi2013} proposed a
radially varying halo shape, characterized by an oblate geometry with $q_z = 0.9$ within
$10\,\mathrm{kpc}$, transitioning to a triaxial form with $(c/a)_\Phi =
0.8$ and $(b/a)_\Phi = 0.9$ at larger radii once the time-dependent
gravitational influence of the LMC (see Section~\ref{sec:challenges}) is
included. This brings the inferred shape back into closer agreement with
cosmological expectations. More recently, \citet{vasilievetal2021} found
that a full joint model of Sagittarius, the LMC, and the Milky Way favors
a still more complex halo, with axis ratios and principal-axis
orientation that vary continuously with radius rather than following any
single fixed triaxial form. Away from Sagittarius specifically,
\citet{bovybahmanyarfritzkallivayalil2016} combined the Pal~5 and GD-1
streams to probe the inner halo and found a shape closer to spherical. This shows that halo flattening inferred from streams need not be
uniform across the Galaxy. Together, these results show
that the discrepancy between idealized and realistic models is large enough to change the inferred halo parameters. Successive triaxial fits to the same
stream have shifted substantially as additional physics, most notably the LMC's time-dependent perturbation, was incorporated. A stream-based halo-shape measurement is only as
reliable as the completeness of the potential model used to obtain it.
\end{sloppypar}

\subsection{Implications for Stream Analysis}
\begin{sloppypar}
The distinction between idealized and realistic models, together with the
sequence of revisions in the Sagittarius-based measurements described
above, has direct consequences for how stream data should be analyzed. A
fit of a single stream with a spherical or axisymmetric potential can
appear to succeed. So can a fit with an early triaxial potential that
omits a relevant physical effect, such as the LMC. In both cases, the
model absorbs real triaxiality, substructure, or time dependence into
biased parameter estimates. A smooth or incomplete model has no other way
to accommodate features it cannot physically produce. This contrast
between idealized and realistic models shows that stellar streams are
effective discriminating probes only when evaluated collectively, as a
whole population, and not individually. Researchers can compare the
properties of multiple streams, spanning different orbits, progenitor
types, and Galactocentric distances, using both classes of models. This
makes it possible to identify which aspects of a best-fit halo are genuine
features of the Milky Way's dark matter distribution and which are
artifacts of an oversimplified or incomplete potential. The result is
tighter constraints on both the macroscopic shape of the halo and the
particle physics that shapes it.
\end{sloppypar}

\section{Future Directions}

\begin{sloppypar}
\textit{Gaia} Data Release 4 is expected in December 2026 and will cover roughly twice the observing baseline of the DR3 catalog \citep{gaiadr3}. The Vera C. Rubin Observatory's LSST
\citep{lsst2019} began in June 2026 to survey the southern sky. These datasets should expand the catalog of known streams well beyond
the nearly one hundred systems currently compiled \citep{mateu2023} and
improve the precision of their kinematics. This will let researchers move from a small sample of well-studied, dynamically modeled systems toward a statistically
significant population, as already validated by all-sky
stream-detection efforts built on algorithms such as STREAMFINDER
\citep{malhanibata2018, ibata2021streamfinder} and by all-sky maps of individual streams \citep{koposov2019orphan}. This
shift from individual stream modeling to statistical ensembles is likely to be one of the most
important observational developments for the field over the coming
decade. Many of the challenges discussed in
Section~\ref{sec:challenges}, which include degeneracies, progenitor sensitivity, and incomplete coverage, are substantially alleviated once the joint statistical frameworks
introduced in Section~\ref{sec:Methodology} can be extended from pairs of streams, as in
\citet{bovybahmanyarfritzkallivayalil2016}, to a large fraction of the
known population.

On the theoretical side, the revision history of the Sagittarius-based
halo-shape measurements described in Section~\ref{sec:Halo Models} is itself a preview of the field's likely trajectory. Successive triaxial fits to the same stream
shifted substantially once the time-dependent influence of the LMC was
incorporated \citep{veraciroaelmi2013, vasilievetal2021}, and it seems likely that
other streams on similarly eccentric or distant orbits are subject to
comparable effects that current models have not yet incorporated. Future work will therefore need to move beyond
fitting each stream to a single fixed potential. A complete approach must model multiple streams simultaneously using potentials that account for radial halo variations, the time-dependent pull of the LMC \citep{erkal2019lmc}, and the gravitational impact of the disk and bar \citep{pearsonpricewhelanjohnston2017}. Achieving this is as much
a statistical and computational challenge as a physical one. It requires
likelihood frameworks capable of combining heterogeneous streams with
different data quality and orbital coverage \citep{sandersonhelmihogg2015,
bonacahogg2018} into a single, internally consistent halo model, together
with the restricted N-body and test-particle simulations needed to
evaluate such models efficiently across a large parameter space
\citep{bovy2015galpy, price-whelangala2020}. Progenitor diversity
\citep{kupper2015} will also need to be inferred simultaneously with the potential rather than fixed, particularly for the many
streams whose progenitor is already fully disrupted.

Perhaps the leading scientific objective, and the one most directly connected
to the broader dark matter problem introduced at the start of this
review, lies in connecting stream signatures to the fundamental nature of
dark matter. Cold dark matter predicts a rich population of small
subhalos, while alternatives such as warm dark matter
\citep{bodeostrikerturok2001} or self-interacting dark matter
\citep{spergelsteinhardt2000} predict fewer, or structurally different,
subhalos. A single stream's gap and spur record, such as that used by
\citet{banik2021} to constrain the dark matter free-streaming scale from
GD-1 alone, already demonstrates that this measurement is possible in
principle; applying the same statistical framework simultaneously across the
much larger stream sample now being assembled \citep{mateu2023,
ibata2021streamfinder} should substantially tighten such constraints,
provided the time-dependent and baryonic complications discussed above
are properly accounted for so that a genuine subhalo signature is not
confused with a feature induced by the Large Magellanic Cloud or the Galactic bar. Such measurements would offer a rare opportunity to probe dark matter
physics on galactic scales with purely gravitational, astrometric data.
They would complement laboratory direct and indirect detection experiments
and cosmological structure-formation constraints \citep{bertone2005, planck2018}. Neither approach measures the low-mass subhalo population of
our own Galaxy directly.
\end{sloppypar}

\section{Conclusion}
\begin{sloppypar}
Dark matter dominates the mass of the Milky Way. The evidence that
established its existence (rotation curves, cluster lensing, and the
cosmic microwave background) says comparatively little about how it is
distributed in detail within any one galaxy, or which of the many proposed
particle candidates it corresponds to (Section~\ref{sec:dmproblem}). Stellar streams have
emerged as one of the most powerful tools for addressing both questions.
Their morphology, whether coherent and continuous or disrupted by gaps,
spurs, and wiggles, records the dynamical influence of the Galactic
potential. This gives a direct observational link between the invisible
structure of the halo and the visible motion of stars. The
five case studies considered in this review, spanning dwarf-galaxy and
globular-cluster progenitors on a diverse range of eccentric, retrograde, inclined, and polar orbits, reflect both the power of this approach
and its present limitations. Even the most intensively studied of these
systems, the Sagittarius stream, has produced a sequence of
increasingly sophisticated but still evolving halo-shape measurements
\citep{lawmajewski2010, degwidrow2012, veraciroaelmi2013,
vasilievetal2021} rather than a single settled answer. The reason is that each successive model included physical processes omitted by earlier studies, including triaxiality, the LMC's 
time-dependent perturbation, and a radially varying shape.

Progress to date reflects the interplay between observational
advances, algorithmic innovation, and theoretical modeling. High-precision
astrometry from \textit{Gaia} \citep{gaia2016, gaiadr2brown2018,
gaiadr3}, systematic detection algorithms
\citep{malhanibata2018} and a growing compilation of the Galaxy's streams \citep{mateu2023} have transformed the observational basis of the field. Simulations and statistical inference
frameworks \citep{bovybahmanyarfritzkallivayalil2016, bonacahogg2018,
sandersonhelmihogg2015} have provided increasingly rigorous tools
for interpreting stream signatures. Challenges remain in halo modeling, particularly regarding parameter degeneracies, incomplete observational coverage, time-dependent perturbations, and progenitor diversity. The
 revision history of the Sagittarius stream is a reminder that these challenges are not
merely theoretical concerns but have already measurably changed
published conclusions about the Milky Way's halo shape. As observational
coverage and theoretical modeling improve together, streams will let us
map the shape of the dark matter halo with growing precision. They will
also constrain the particle nature of dark matter through small-scale
substructure. Streams around other galaxies will extend these tests to a
much broader population of halos across the Universe. Taken together, these advances point toward a halo that is measured
stream by stream and then tested across a whole population.
\end{sloppypar}

\section*{Acknowledgments}
\begin{sloppypar}
This review includes two publicly released data products from the
European Space Agency (ESA) mission \textit{Gaia}
(\url{https://www.cosmos.esa.int/gaia}), shown here as
Figures~\ref{fig:sgr-gaia} and \ref{fig:gd1-gaia} and processed by the
\textit{Gaia} Data Processing and Analysis Consortium (DPAC,
\url{https://www.cosmos.esa.int/web/gaia/dpac/consortium}). Funding for
the DPAC has been provided by national institutions, in particular the
institutions participating in the \textit{Gaia} Multilateral Agreement.
These figures are used here unmodified under a CC~BY-SA~3.0~IGO license,
with credit lines given in the corresponding figure captions.

The original schematic figures in this review (Figures~\ref{fig:rotcurve},
\ref{fig:streamformation}, and \ref{fig:gapspur}) were produced using
NumPy \citep{harris2020numpy} and Matplotlib \citep{hunter2007matplotlib}.
Code development and figure generation were carried out using Google
Colaboratory (\url{https://colab.research.google.com}).

This research has made use of the Astrophysics Data System, funded by
NASA under Cooperative Agreement 80NSSC21M0056.

We also acknowledge the use of Claude to assist with refining manuscript language, grammar, punctuation, and phrasing. Following AI assistance, all text, technical descriptions, citations, and scientific interpretations were meticulously reviewed and re-verified by the authors.
\end{sloppypar}

\enlargethispage{0.5\baselineskip}
\bibliographystyle{plainnat}
\bibliography{references}

\end{document}